\documentclass[10pt,journal] {IEEEtran}
\usepackage{graphicx} % Required for inserting images
\usepackage{cite}
\usepackage{subcaption}
\usepackage{amsmath}
\usepackage{authblk}
\usepackage{enumitem}
\begin{document}
\title{ Quantum Key Distribution Routing Protocol in Quantum Networks: Overview and Challenges
% \thanks{Funding Details}
}

\author{
Pankaj Kumar, {\em Student Member, IEEE}, Neel Kanth Kundu, {\em Member, IEEE}, and Binayak Kar, {\em Member, IEEE}
\thanks{
P. Kumar and B. Kar are with Department of Computer Science and Information Engineering, National Taiwan University of Science and Technology, Taiwan (e-mail: pnkazaayan@gmail.com, bkar@mail.ntust.edu.tw).

Neel Kanth Kundu is with the Centre for Applied Research in Electronics (CARE) and Bharti School of Telecommunication Technology and Management (BSTTM), Indian Institute of Technology (IIT) Delhi, New Delhi-110016, India (e-mail: neelkanth@iitd.ac.in)
}
}

% \author{
% \IEEEauthorblockN{Pankaj Kumar$^{1}$, Neel Kanth Kundu$^{2}$ and Binayak Kar$^{1}$} 

% \IEEEauthorblockA{$^{1}$Department of Computer Science and Information Engineering, National Taiwan University of Science and Technology, Taiwan}

% \IEEEauthorblockA{ $^{2}$Department of Electrical and Electronic Engineering, University of Melbourne, Australia}

% Email: \texttt{pnkazaayan@gmail.com, neelkanth.kundu@unimelb.edu.au, bkar@mail.ntust.edu.tw}
% }

\maketitle

\begin{abstract}
    The use of quantum cryptography in everyday applications has gained attention in both industrial and academic fields. Due to advancements in quantum electronics, practical quantum devices are already available in the market, and ready for wider use. Quantum Key Distribution (QKD) is a crucial aspect of quantum cryptography, which involves generating and distributing symmetric cryptographic keys between geographically separated users using principles of quantum physics. Many successful QKD networks have been established to test different solutions. The objective of this paper is to delve into the potential of utilizing established routing design techniques in the context of quantum key distribution, a field distinguished by its unique properties rooted in the principles of quantum mechanics. However, the implementation of these techniques poses substantial challenges, including quantum memory decoherence, key rate generation, latency delays, inherent noise in quantum systems, limited communication ranges, and the necessity for highly specialized hardware. This paper conducts an in-depth examination of essential research pertaining to the design methodologies for quantum key distribution. It also explores the fundamental aspects of quantum routing and the associated properties inherent to quantum QKD. This paper elucidates the necessary steps for constructing efficient and resilient QKD networks. In summarizing the techniques relevant to QKD networking and routing, including their underlying principles, protocols, and challenges, this paper sheds light on potential applications and delineates future research directions in this burgeoning field.
\end{abstract}

\section{introduction} \label{1}
Creating secure cryptographic keys over untrusted networks is crucial \cite{maurer1993secret}. While public key infrastructure, based on complex mathematical problems and assumptions about the computational capability of eavesdroppers, is widely used, it falls into the category of theoretically vulnerable computational security solutions. With the continual advancement of computational power and the emergence of quantum computing algorithms capable of solving complex mathematical problems in polynomial time, such as Shor's factoring algorithm \cite{shor1994algorithms}, new challenges and opportunities are on the horizon. Quantum Key Distribution (QKD) operates on the principles of quantum information theory and allows for the creation of information-theoretic secure cryptographic keys \cite{shor2000simple}.

To ensure the integrity and security of data packets traversing through the QKD network, it is imperative to encrypt them using quantum secure keys. This study emphasizes the significance of extending this security paradigm to the routing packets themselves, highlighting the critical need to safeguard routing data with quantum secure keys. In essence, QKD networks distinguish themselves by their heavy reliance on quantum secure keys as a fundamental element in their communication architecture \cite{nejatollahi2019post}. Traditional routing protocols, designed without the consideration of this unique feature of QKD networks, pose a substantial challenge when directly applied within QKD infrastructures. Such a direct application invariably results in the suboptimal utilization of quantum secure keys. This underutilization is a highly undesirable outcome, given the intrinsic value and scarcity of quantum secure keys within the QKD network ecosystem. Consequently, the development and implementation of a specialized routing protocol capable of achieving efficient and high utilization of quantum secure keys emerge as critical prerequisites for the practical deployment and success of QKD networks \cite{zou2020collaborative}.

QKD networks exhibit substantial distinctions in comparison to conventional telecommunication networks, primarily stemming from the unique characteristics of QKD links and the intricacies of network organization. QKD encounters several technical limitations. Such as low-key generation rates and communication distance due to factors like photon detection efficiency, transmission losses, and noise levels. The absence of practical quantum repeaters poses a significant hurdle, as they are essential for extending QKD's range beyond its current limitations. QKD systems necessitate precise routing strategies due to the coexistence of classical and quantum channels within quantum links. Consequently, contemporary network architectures employ a hop-by-hop approach for secure key distribution to navigate these complexities effectively \cite{wang2021quantum}. QKD Networks comprise several quantum repeaters and trusted nodes that facilitate communication between Alice and Bob. These components play a crucial role in enabling significantly extended communication distances by reducing signal loss through shorter hops between nodes. Additionally, trusted nodes perform measurements between parties, enhancing security. Another advantage of QKD networks is the existence of multiple paths between Alice and Bob, leading to increased key generation rates as key bits are distilled on each connecting path in an additive manner \cite{cao2022evolution}. 

% This paper focuses on QKD networking, network organization, routing and signaling protocols, and software-defined QKD networking techniques.

This paper focuses on the networking aspects of QKD with an emphasis on QKD network organization, routing and signaling protocols, and software-defined QKD networks. This paper aims to provide the readers with an insight into QKD routing design from an engineering perspective. It is important to incorporate the interdependence of conventional cryptography and QKD for the successful design of QKD routing \cite{schoute2016shortcuts}.
% The interdependence between conventional cryptography and QKD must be carefully evaluated and considered for the successful design of QKD routing \cite{schoute2016shortcuts}.
However, this interdependence remains poorly understood despite its fundamental importance, leading to significant open issues in QKD network design. The primary objective of this article is to provide an understanding of strategies for designing QKD routing techniques in quantum networks, empowering readers to:
% The primary aim of this article is to offer insight into QKD routing design techniques in quantum networks, enabling readers to:

    \begin{itemize}
        \item { Identify effective QKD routing protocol design methods within a quantum network in order to comprehend the information flow within the network.}
        
        % {Recognize the efficient design techniques of QKD routing protocols in a quantum network to comprehend information flow within the network.}
        \item {Analyze quantum networks from an engineering standpoint and gain proficiency in understanding their operational modes, implementation approaches, currently available solutions, and techniques for simulating quantum cryptographic networks.}
        \item{Provide a high-level perspective on the design of QKD networks with a focus on applied quantum cryptography techniques for future communication networks.}
        % \item{It provides a high-level view of QKD network design, and technique in the field of applied quantum cryptography for future communication networks.}
    \end{itemize}

\begin{figure}[h]
    \centering
    \includegraphics[width=0.7\linewidth]{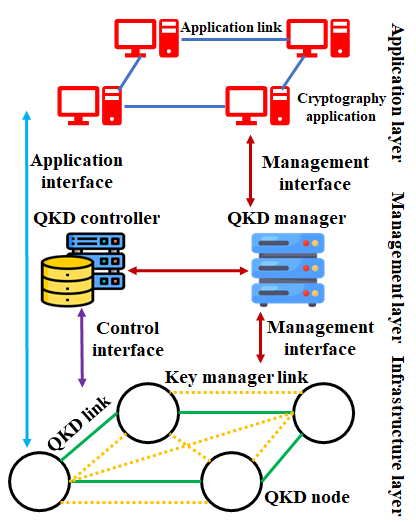}
    \caption{QKD network architecture}
    \label{fig:1}
\end{figure}

The rest of the paper is structured as follows: In Section \ref{1}, we provide an overview of the QKD network. Section \ref{2} presents QKD network architecture and elements. The QKD routing design techniques are discussed in Section \ref{3} and future research direction in Section \ref{4}. Lastly, we summarize our findings and conclusions in Section \ref{5}.

\section{QKD NETWORK ARCHITECTURE AND ELEMENTS} \label{2}
\subsection{QKD Networks Architecture:} The QKD network cannot exist independently of the classical network, as it necessitates an authenticated classical network and various secure cryptographic functions within the classical network. The architecture of networks can vary significantly, characterized by differing numbers of layers based on their unique definitions and intended applications. The ITU-T Y.3800 recommendation \cite{recommendation20193800} provides comprehensive insights into the conceptual structures of both QKD networks and user networks. In the context of network architectures that facilitate QKD, a comprehensive three-layer design has been developed from a holistic perspective. This design draws inspiration from the established six-layer network architecture. As illustrated in Figure \ref{fig:1}, this network architecture comprises three distinct logical layers.

\emph{Infrastructure Layer:} In Figure \ref{fig:1}, the infrastructure layer of the QKD network comprises a range of essential physical components necessary for QKD networking. These components are strategically situated within secure and dependable nodes, referred to as QKD nodes, with the primary objective of safeguarding against potential physical attacks. QKD nodes have the capability to establish connections in pairs through either optical fiber or free-space links, facilitating the generation of secret keys in the form of symmetric random bit sequences. Notably, these secret keys, being classical bit sequences, are securely stored within the QKD nodes. Furthermore, each QKD node diligently maintains comprehensive secret-key parameters, including identifiers, sizes, transmission rates, key types, as well as physical device identification, and timestamp records, which are utilized for both the generation and storage of secret keys \cite{wang2021quantum}. Additionally, QKD nodes retain pertinent information concerning the link parameters, encompassing details such as link length, type, and the error rate associated with quantum channels.

\emph{Control and Management Layer:} In Figure \ref{fig:1}, the higher-level layer consists of the QKD network controller and manager, as described in \cite{recommendation20193800}. The QKD network controller is tasked with the responsibilities of activating, deactivating, and calibrating all QKD nodes within the network. On the other hand, the role of the QKD network manager involves the comprehensive oversight and maintenance of the entire QKD network. This includes the continuous monitoring of the status of all QKD nodes and links, along with real-time data acquisition of secret-key parameters and link parameters from these nodes. Additionally, the QKD network manager is responsible for supervising the operations of the QKD network controller. Statistical information obtained through monitoring and management is regularly collected, recorded, and updated within a designated database to facilitate data analysis and historical tracking. It's important to highlight that the actual secret keys stored within the QKD nodes are securely maintained in physically isolated locations. This design ensures that the secret keys remain inaccessible to both the QKD network controller and the network manager. This robust security measure safeguards the secrecy of the keys, even with the introduction of the control and management layer \cite{amer2020efficient}.

\emph{Application Layer:} The highest layer depicted in Figure 1 is the QKD application layer, which encompasses cryptographic applications that are integral to users of the QKD network. This process involves cryptographic applications communicating their security requirements, such as specific secret key specifications, to the QKD network manager. Subsequently, the manager undertakes the task of verifying the availability of the requested secret keys. If the requisite keys are indeed available, the manager proceeds to instruct the QKD network controller to furnish these keys to the applications for the purpose of data encryption. Upon receiving the supplied secret keys, each application assumes responsibility for its management and usage. It's important to note that the capacity of a QKD network to accommodate users is contingent upon the availability of resources and the particular requirements of the users themselves \cite{yang2018quantum}.

\subsection{QKD Network Elements:} Considering the architecture of QKD networks, the corresponding QKD network components can be delineated as follows:

\emph{QKD Nodes:} QKD nodes are essential components within a Quantum Key Distribution network. These nodes play a crucial role in the establishment of secure cryptographic keys based on the principles of quantum mechanics. QKD nodes are responsible for generating cryptographic keys, often referred to as quantum keys. These keys are created using the unique properties of quantum mechanics, such as superposition and entanglement of photons, in order to ensure their security against eavesdropping. These nodes are designed to be physically secure to protect against tampering and attacks. Once quantum keys are generated, QKD nodes are responsible for securely distributing these keys to users or applications that require them for encryption and decryption processes \cite{maurer1993secret}.

\emph{QKD Network Controller:} The QKD network controller shown in Figure \ref{fig:1} is a centralized server responsible for managing QKD nodes within a QKD network infrastructure. Its functions include activating, deactivating, and calibrating QKD nodes. The QKD network controller assumes critical responsibilities in network management. These encompass essential network control functions such as QKD connection management, which includes tasks like node access and authentication. The controller also manages routing control, enabling secret-key relay and facilitating failure recovery within the network. Additionally, it plays a pivotal role in Quality of Service (QoS) control, encompassing aspects like customization and ensuring end-to-end QoS assurance.

\emph{Cryptographic Application:} The top layer in Figure \ref{fig:1} represents a cryptographic application, which is essentially a user with particular security requirements. These requirements typically involve requests for cryptographic keys, specifying details like key size, rate, and key refresh rate. Importantly, a cryptographic application is usually co-located in the same physical location as a QKD node to receive the secret keys. This proximity is essential for the secure exchange of keys.

\begin{figure}
    \centering
    \includegraphics[width=0.7\linewidth]{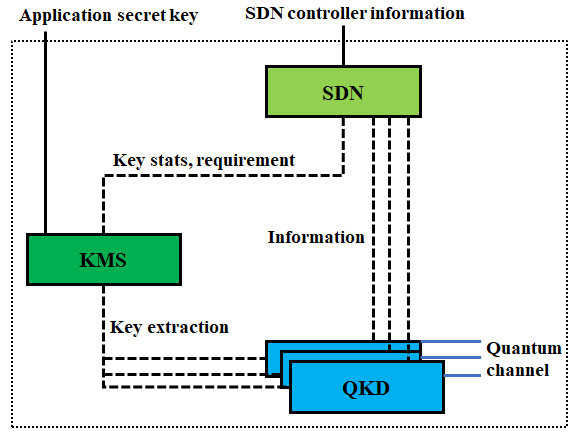}
    \caption{SDN-based QKD}
    \label{fig:2}
\end{figure}
\section{Routing Design Techniques for QKD Networks}  \label{3}
Over the recent years, substantial strides have been taken to tackle the technical hurdles associated with practical QKD networking. This section delves into an extensive examination of the design methodologies introduced for QKD networks, encompassing Software-Defined Networking (SDN), routing strategies, overlay solutions, and switch-centric QKD approaches.

\subsubsection{SDN-based QKD} The QKD scheme proposed in \cite{aguado2019engineering} aims to make QKD systems compatible with SDN principles. This involves creating a model for QKD systems that aligns them with traditional network elements like routers and switches. This model necessitates careful consideration of which QKD capabilities and parameters are relevant from a network management perspective. For instance, the intricate nature of QKD systems, with multiple complex hardware and software components, needs to be abstracted to simplify interaction with the SDN controller. Additionally, sensitive data like internal information and actual key material must be kept confidential to ensure the security of the QKD network. This model serves several purposes: it promotes the adoption of standard interfaces, reducing time-to-market for solutions that previously relied on custom ad-hoc implementations. It leverages SDN technologies for optical networks to improve quantum channel spectrum allocation. Furthermore, centralizing QKD network management enhances the optimization of key generation and utilization while maintaining a centralized database of active links and consumer applications.

The network involves aggregating and abstracting QKD systems within the same secure area as logical network elements. This is represented in the node architecture, where various QKD systems are controlled within the Software-Defined QKD (SD-QKD) node shown in Figure \ref{fig:2}. The SD-QKD node model comprises four essential components:

\begin{enumerate}[label=(\alph*)]
\item {\textit{Interfaces:} These encompass all the QKD systems that are consolidated within an SD-QKD node. These interfaces facilitate the interaction between the node and the QKD systems.}

\item {\textit{Applications:} Applications can be either internal or external entities that make use of keys obtained from the node's Key Management System (KMS). These keys are crucial for secure communications and are accessed by various applications.}

\item {\textit{Links:} These are akin to key connections between two SD-QKD nodes. These links can take on two forms: physical connections via a quantum channel or virtual connections established through trusted relay mechanisms. They play a pivotal role in securing the communication between nodes.}

\item {\textit{Capabilities and additional information:} This component encompasses supplementary data, including identifiers and geographical locations. It also provides support for various operations required for the effective functioning of the SDQKD node. These capabilities and additional information are crucial for managing and optimizing the quantum key distribution network.}
\end{enumerate}
These physical QKD systems create quantum channels and generate quantum secure keys, which are managed by KMS. Finally, an SDN agent plays a crucial role in gathering data from the QKD node, establishing communication with the SDN controller, and implementing configuration updates as requested by the QKD controller. This requires the agent to coordinate with other elements within the node, such as the KMS and the QKD systems.

\begin{figure}
    \centering
   \includegraphics[width=0.7\linewidth]{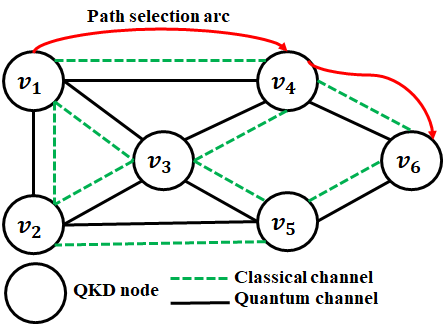}
    \caption{Routing and key pool-based QKD}
    \label{fig:3}
\end{figure}
\subsubsection{Routing and key pool-based QKD} 
% Chao Yang et al. \cite{yang2018quantum} describe the QKD network routing model based on graph theory. 
A graph theoretic-based QKD network routing model was described in \cite{yang2018quantum}. In this paper, it is assumed that the QKD network comprises many independent point-to-point (P2P) QKD systems and each P2P QKD system continuously generates a secret key for two adjacent end nodes. In the QKD network, all nodes are interconnected in the classical networking perspective, however, quantum channels are only present between specific nodes. Additionally, each node has the capability to establish a neighbor relationship solely with nodes directly linked by quantum channels. Figure \ref{fig:3} shows a topology of the QKD network. 

Every node within the network possesses two distinct communication channels: a classical channel and a quantum channel. The classical channel functions as a virtual connection within the classical network and is represented by a dotted line in Figure \ref{fig:3}. Its primary purpose is to facilitate the transmission of essential network control information, encrypted data, and similar content. On the other hand, the quantum channel is formed through the optical fibers linking two neighboring nodes, depicted as a solid line in Figure \ref{fig:3}. The primary role of the quantum channel is to facilitate the creation of secret keys for each QKD system. The key generated by the P2P QKD system is the local key. In the QKD network, each P2P QKD system operates autonomously and in a continuous manner to generate local keys for the two nodes at either end of the communication link. These local keys are subsequently stored in dedicated secret key pools, referred to as local key pools. The concept of a local key serves as the fundamental building block for establishing secure keys for users within a network. These locally usable keys are securely stored in what we term a residual local key pool. In specific scenarios \cite{diadamo2022packet}, where a quantum channel is unavailable, the generation of a local key between two distant end nodes becomes challenging. In such circumstances, these two end nodes have no choice but to establish shared keys through a relay process with the assistance of multiple intermediate nodes. The keys generated through this relaying process, which we refer to as global keys, necessitate the utilization of a specific quantity of local key resources for their creation and maintenance. This approach ensures that secure communication is maintained even when faced with the absence of a direct quantum channel, allowing for the establishment of shared keys through a collaborative network effort. This global key serves as the actual secret key employed for secure communication among users, encompassing activities like voice calls, video chats, and file transfers. Consequently, the global key can be established between nodes that are either far apart or adjacent to each other. To ensure secure communication, when two neighboring nodes require a global key, they can directly utilize the local keys they share as the global key. In cases where nodes are not adjacent and still require a global key, they must acquire it through a process involving multiple intermediary nodes facilitating the relay of keys. The procedure by which nodes acquire a global key is commonly referred to as key exchange. During the exchange of the global key from its source to its intended destination, it often traverses multiple intermediate QKD links. This intermediate QKD link collection is collectively called the relaying path.

\begin{figure}
    \centering
    \includegraphics[width=0.8\linewidth]{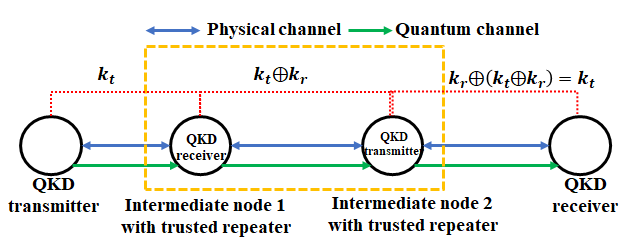}
    \caption{Trusted repeater-based QKD}
    \label{fig:4}
\end{figure}
\subsubsection{Trusted repeater-based QKD} Point-to-point QKD encounters limitations arising from various physical-layer impediments. These include issues like scattering and the attenuation of quantum signals during their passage through quantum channels. Additionally, it is worth noting that quantum states of light cannot be amplified. This is due to the fundamental requirement of measuring and cloning a quantum state of light, which, as outlined in the quantum no-cloning theorem \cite{mehic2020quantum}, is inherently impossible. Consequently, the establishment of long-distance end-to-end QKD connections necessitates the integration of repeater technologies. Quantum repeaters, defined as devices capable of forwarding qubits without engaging in measurement or cloning processes, continue to pose challenges in practical QKD network implementations. The trusted repeater technique has emerged as a pragmatic solution to address this challenge, commonly employed in real-world QKD networks. This approach leverages intermediate nodes equipped with trusted repeaters, facilitating the gradual transmission of secret keys along the QKD path, originating from the source node and culminating at the destination node. In this process, secret keys undergo decryption and subsequent re-encryption. This operation is executed using an information-theoretically secure One-Time Pad (OTP) cryptographic algorithm \cite{shor1994algorithms} at each intermediate node. 

Figure \ref{fig:4} shows long-distance end-to-end QKD based on trusted repeaters. We've strategically positioned two intermediate nodes equipped with trusted repeaters for the source and destination nodes. This setup facilitates secure communication within a QKD framework \cite{yang2018quantum}. The QKD sender is linked to the QKD receiver through a combination of Quantum channels and public channels. The key objective here is to establish three distinct secret keys, each serving specific communication segments: the source node to the intermediate node 1, intermediate node 1 to intermediate node 2, and intermediate node 2 to the destination node. It's important to note that all three secret keys share identical key sizes, ensuring uniform cryptographic strength across the communication path. The process unfolds in a series of meticulously orchestrated steps:

\begin{enumerate}[label=(\alph*)]
    \item {\textit{Source node to Intermediate node 1:} In the first step, a secret key is generated securely between the source node and intermediate node 1. This step is vital as it forms the foundation for secure communication between the source and the initial intermediate node.}

    \item {\textit{Intermediate node 1 to Intermediate node 2:} The second step involves the creation of a secret key, establishing a secure link between Intermediate Node 1 and Intermediate Node 2. This intermediary key bridges the gap between the two consecutive nodes, ensuring the confidentiality and integrity of the data in transit.}

    \item {\textit{Intermediate Node 2 to Destination node:} Finally, in the third step, a secret key is generated to establish secure communication between intermediate node 2 and the destination node. This last link in the chain ensures the end-to-end security of the communication, safeguarding data as it reaches its intended destination.}
\end{enumerate}
This meticulous three-step process, enabled by the strategic placement of intermediate nodes with trusted repeaters, allows for the generation of three distinct secret keys, each with uniform key sizes, to secure the communication channels between the source node and the destination node. This approach ensures the confidentiality and integrity of data throughout its paths.

\begin{figure}
    \centering
     \includegraphics[width=0.7\linewidth]{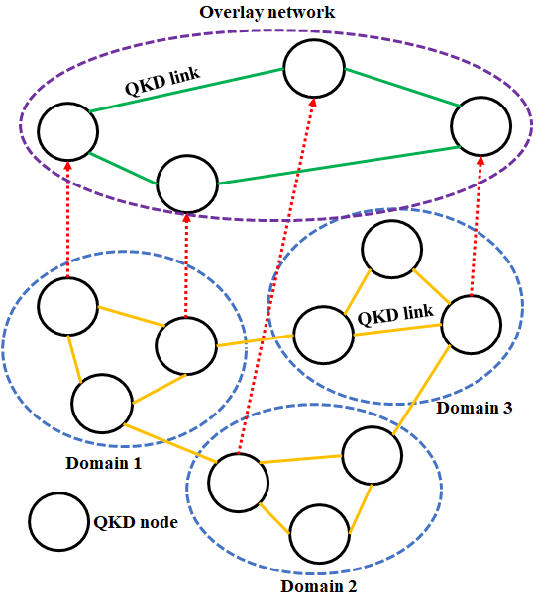}
    \caption{Overlay-based QKD}
    \label{fig:5}
\end{figure}
\subsubsection{Overlay-based QKD} The previously discussed QKD network types deal with the arrangement of quantum channels, whereas the QKD overlay network type is focused on the realization of public channels. The primary objective of the overlay network is to establish a higher-level network structure to enhance QoS and make efficient use of resources from lower-level networks. In pursuing this goal, the overlay network seeks to operate independently of predefined paths set by Internet Service Providers (ISPs). Key characteristics of the overlay network approach include the ability to discover alternative routes that offer superior service quality and rapid rerouting in case of interruptions, as well as the utilization of multipath communications \cite{kar2023routing}. Employing multipath connections is a frequently recommended solution for enhancing network performance, encompassing safeguards against network failures, distribution of network loads, implementation of high bandwidth capabilities, and selection of low-delay pathways \cite{amer2020efficient}. 

The use of external routing protocols like the Border Gateway Protocol (BGP) for routing between network domains is acknowledged to result in sluggish response times and delayed recovery from network disruptions. This is primarily due to the time it takes to gather information about network link interruptions or congestion, as well as the typically configured minimum route advertisement interval timer in BGP, which often operates on a timescale of minutes. Consequently, the process of achieving a consistent network view following a link outage can extend to tens of minutes, a substantial timeframe that poses challenges for network applications. Moreover, BGP's propagation of only a single route makes it challenging for network nodes to detect alternative routes in various scenarios \cite{kar2023routing}.

To address these challenges, the overlay network adopts a peer-to-peer approach, facilitating connections between nodes in diverse domains and enabling the utilization of alternative pathways through traffic encapsulation within the lower network. As a packet traverses the path, intermediate nodes receive it, unbox the packet, scrutinize the recipient's IP address, re-encapsulate the packet, and then transmit it onward to potential network nodes in other domains. Essentially, it operates as a step-by-step process commonly implemented in QKD networking shown in Figure \ref{fig:5}. Leveraging the encapsulation principle, overlay nodes autonomously gauge link conditions and can promptly respond to congestion by diverting traffic towards less congested routes. Overlay networks introduce novel capabilities that are challenging to achieve within lower-level networks. The appeal of the overlay QKD approach lies in its ability to bypass "untrusted" nodes and swiftly reconfigure routes when trust in certain nodes becomes compromised or when multipath communication is essential \cite{schoute2016shortcuts}.

\section{Future Research Directions}  \label{4}

\emph{1) Quantum repeaters:} Quantum repeaters are essential for extending the reach of QKD networks beyond their current limitations. One of the primary limitations of traditional QKD is the distance over which quantum keys can be reliably transmitted. Quantum signals are susceptible to attenuation and loss over long distances. Quantum repeaters address this limitation by effectively "renewing" and forwarding quantum information, allowing secure communication over much greater distances. In a QKD network, quantum signals can degrade significantly as they pass through optical fibers or other communication channels. Quantum repeaters mitigate this loss by actively enhancing and amplifying the quantum signals while preserving their quantum properties. This ensures that the quantum keys remain secure and usable over extended distances. Future work should focus on the development of efficient and practical quantum repeater protocols and technologies, which can enable secure communication over longer distances.

\emph{2) High-rate QKD:} Traditional QKD systems have been limited by relatively low-key generation rates. These systems could not match the data transfer rates commonly used in modern networks and applications. This limitation made QKD less practical for high-speed data transmission needs, such as those in data centers, financial institutions, and critical infrastructure sectors. Low-key generation rates have been primarily attributed to the nature of quantum signals and the stringent security requirements of QKD. Quantum signals, typically carried by individual photons, have inherently low signal-to-noise ratios, leading to slow key generation processes. Moreover, the security of QKD relies on the need for extremely long and random keys, further impeding the rate at which keys can be produced.

\emph{3) Quantum network topologies:} The choice of quantum network topology is a critical decision that can significantly impact the performance, security, and scalability of a Quantum Key Distribution network. A well-designed topology considers the unique properties of quantum communication while aligning with the network's goals and applications \cite{mehic2020quantum}. It plays a central role in ensuring the reliable and secure exchange of quantum keys across distributed nodes and users.

\emph{4) Quantum cryptography protocol:} A Quantum network protocol in QKD networks is a set of rules and procedures governing the exchange of quantum information, primarily quantum keys, among multiple network nodes in a secure and efficient manner. These protocols are designed to harness the principles of quantum mechanics to establish cryptographic keys with provable security properties, such as resistance to eavesdropping attacks. Quantum network protocols dictate how quantum states, often in the form of entangled particles or qubits, are generated, transmitted, and measured between different nodes within the network \cite{yang2018quantum}. They also include procedures for error correction, privacy amplification, and key distillation to ensure the final keys are both secure and reliable.

\emph{5) Practical implementations and cost reduction:} Practical implementations and cost reduction are pivotal considerations in the development of QKD networks. To make quantum communication technology feasible for real-world applications, researchers and engineers are focused on creating more compact, robust, and affordable QKD hardware and infrastructure. Practical implementations involve designing QKD systems that are easy to install and maintain, even in resource-constrained environments. Furthermore, cost reduction efforts seek to bring down the expenses associated with QKD networks, including the cost of specialized quantum components, integration into existing networks, and overall operational costs. Achieving practicality and cost-effectiveness is critical to making QKD accessible to a broader range of users and industries, ultimately driving the adoption of quantum-secure communication solutions in sectors such as finance, healthcare, and telecommunications.

\section{Conclusion}  \label{5}
QKD networks offer the promise of long-term data security and robust protection for a wide array of applications. However, they are not without their share of unresolved challenges and open questions. This paper aims to provide a comprehensive overview, delving into past accomplishments while offering insights into the future of QKD networks. This work categorizes various options for implementing QKD networks and reviews their development, encompassing trusted repeaters, SDN architectures, and routing-based key pools within QKD networks \cite{waxman1988routing}. Furthermore, the paper presents a detailed examination of the general architecture of QKD networks, highlighting the constituent elements, interfaces, and protocols involved. This paper outlines a promising set of research directions that point toward the initial steps in shaping the quantum internet. Achieving comprehensive progress in QKD network development will necessitate collaborative efforts across multiple domains, including physics, computer science, security, and communications.
\bibliographystyle{IEEEtran}
\bibliography{cite.bib}

\begin{IEEEbiographynophoto}{Pankaj Kumar}
 earned his M.Tech. degree in computer science and engineering from the Indian Institute of Technology (ISM) in Dhanbad, India. He is currently pursuing a Ph.D. degree in computer science and information engineering (CSIE) at National Taiwan University of Science and Technology (NTUST), Taiwan. His research focuses on network design, quantum algorithms, quantum information theory, quantum communications, and QKD networks.
 \end{IEEEbiographynophoto}

\begin{IEEEbiographynophoto}{Neel Kanth Kundu}
is an Assistant Professor at the Centre for Applied Research in Electronics (CARE) and Bharti School of Telecommunication Technology and Management, Indian Institute of Technology Delhi.  He received his Ph.D. degree in electronic and computer engineering with a concentration in scientific computation from The Hong Kong University of Science and Technology, Clear Water Bay, Hong Kong in 2022. He was a postdoctoral research fellow with the Department of Electronic and Electrical Engineering at the University of Melbourne,
Australia from February 2023 to October 2023. His research
interests include signal processing for 6G wireless communications, quantum communications, and quantum information processing.
\end{IEEEbiographynophoto}
 
\begin{IEEEbiographynophoto}{Binayak Kar}
is an Assistant Professor of computer science and information engineering at National Taiwan University of Science and Technology (NTUST), Taiwan. He received his Ph.D. degree in computer science and information engineering from the National Central University (NCU), Taiwan, in 2018. He was a post-doctoral research fellow in computer science at National Chiao Tung University (NCTU), Taiwan, from 2018 to 2019. His research interests include network softwarization, cloud/edge/fog computing, optimization, queueing theory, machine learning, cyber security, and quantum computing.
\end{IEEEbiographynophoto}

\end{document}